\documentclass{article}

\usepackage{arxiv}
\usepackage[utf8]{inputenc} 
\usepackage[T1]{fontenc}    
\usepackage{hyperref}       
\usepackage{url}            
\usepackage{booktabs}       
\usepackage{amsfonts}       
\usepackage{nicefrac}       
\usepackage{microtype}      
\usepackage{lipsum}
\usepackage{graphicx}
\graphicspath{ {./images/} }
\usepackage{subcaption}
\usepackage{adjustbox}
\usepackage{amsmath}
\usepackage{multirow}
\usepackage{makecell}

\begin{document}

\title{TPMM-DPO: Trajectory-aware Preference-guided Model Merging for Iterative Direct Preference Optimization}
\author{
Lingling Fu$^{1}$ \quad Yongfu Xue$^{2}$\\
$^{1}$Guangxi University\\
\texttt{2313593009@st.gxu.edu.cn, xueyongfu@outlook.com}
}
	
	\maketitle
	
	\begin{abstract}

        Direct Preference Optimization (DPO) has been widely adopted for large language model alignment due to its simple training procedure and lack of an explicit reward model. However, in iterative DPO, when the policy model from the previous iteration is repeatedly used as the reference model for subsequent rounds, noise in preference data and errors in the reference model accumulate over time. This accumulation can lead to late-stage over-optimization, performance fluctuations, and degraded generalization.
        To address these issues, we propose TPMM-DPO, a  trajectory-aware preference-guided model merging method. The method treats the sequence of policy models generated during iterative DPO as an optimization trajectory and adaptively integrates them using learned fusion weights, thereby constructing a smoother and more robust reference model. In contrast to conventional iterative DPO, which relies solely on a single previous model, TPMM-DPO effectively mitigates error accumulation induced by noisy preferences and improves training stability.
        Experimental results show that standard iterative DPO often suffers from performance degradation in the middle and later stages of training, whereas TPMM-DPO consistently improves generation quality and achieves higher win rates and reward scores on both in-domain and out-of-domain evaluations. Further ablation studies and robustness analyses demonstrate that, compared with simple averaging, learnable-weight fusion more effectively alleviates late-stage performance degradation caused by noisy preferences.
	\end{abstract}


	\section{Introduction}

    Direct Preference Optimization (DPO)~\cite{rafailov2023direct} significantly simplifies the training pipeline of reinforcement learning from human feedback (RLHF)~\cite{ouyang2022training,glaese2022improving} compared with online reinforcement learning methods such as PPO~\cite{schulman2017proximal}. Given offline preference data, DPO directly optimizes the policy model by converting the preference that a chosen response should be preferred over a rejected response into a supervised objective over log-probability differences. This avoids the complexity and instability introduced by explicit reward-model training and online sampling.

Despite its simplicity and relatively stable optimization behavior, DPO remains highly dependent on the quality of preference data. In practice, preference datasets often contain label noise, annotation inconsistency, and distribution shift. These issues may cause large language models~\cite{achiam2023gpt,ouyang2022training} to learn incorrect preference signals and therefore undermine alignment performance~\cite{chowdhury2024provably}. Moreover, because DPO does not include an online exploration mechanism, it has limited ability to correct out-of-distribution failures, making the problem more severe in noisy settings.

Iterative DPO has been introduced to reduce distribution shift by repeatedly generating candidate responses with the current policy and constructing new preference data for the next round of optimization. In the standard setting, however, the policy obtained in the previous round is directly used as the reference model for the next round. When preference data are noisy, incorrect preference signals can be inherited and amplified across iterations. As a result, the reference model may gradually drift away from the true preference distribution, causing training oscillation and performance degradation.

In this paper, we propose TPMM-DPO, a learnable model merging method based on historical policy trajectories. Our goal is to alleviate noise accumulation and late-stage over-optimization in iterative DPO. Instead of using only the most recent policy as the next reference model, we treat the policies produced during iterative optimization as a trajectory and adaptively merge policies from different stages in parameter space. The merged model is used as the reference model for the next DPO round. By learning the contribution of each historical policy, TPMM-DPO exploits complementary strengths across training stages, builds a smoother reference distribution, and reduces instability caused by noisy preferences. Experimental results show that TPMM-DPO improves both training stability and final performance under noisy preference data.

\begin{figure}[htbp]
    \centering
    \includegraphics[width=0.8\linewidth]{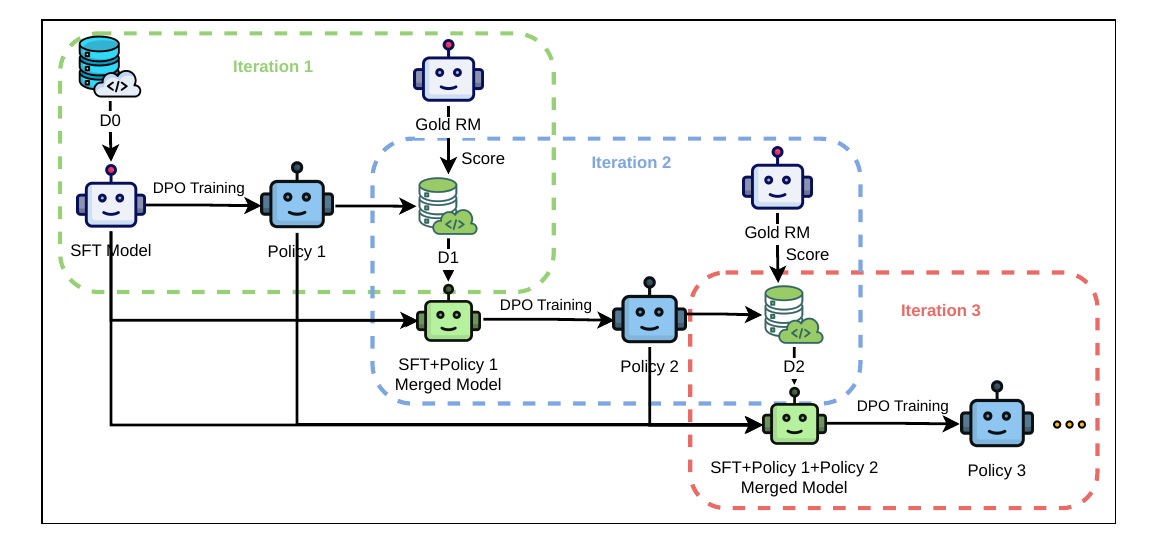}
    \caption{Overview of the proposed TPMM-DPO training framework.}
    \label{fig:mergedpo}
\end{figure}

\section{Background}
\subsection{RLHF}
Reinforcement Learning from Human Feedback (RLHF)~\cite{ouyang2022training,glaese2022improving} is a central paradigm for aligning large language models. Given a prompt $x$, a model generates multiple candidate responses, and human annotators compare them in pairs to provide preference data. A conventional RLHF pipeline usually consists of three stages. First, supervised fine-tuning (SFT)~\cite{muennighoff2025s1,ye2025limo} is used to obtain an initial aligned model. Second, a reward model is trained from preference data to approximate human preferences. Finally, a reinforcement learning algorithm such as PPO is used to optimize the policy model by maximizing the reward-model output while applying a KL constraint to prevent the policy from drifting too far from the initial distribution.

Although RLHF has achieved strong empirical performance, its training pipeline is complex and computationally expensive. Errors in the reward model can also be amplified during reinforcement learning, leading to reward hacking or over-optimization. These challenges make RLHF difficult to implement robustly in practice.

\subsection{DPO}

Rafailov et al.~\cite{rafailov2023direct} proposed Direct Preference Optimization (DPO) as a simpler alternative to RLHF. DPO does not explicitly train a reward model and does not rely on policy-gradient optimization. Instead, it directly optimizes the policy using offline preference pairs. For a prompt $x$ and a preference pair $(y_w,y_l)$, DPO maximizes the log-probability advantage of the preferred response over the rejected response under a fixed reference model. The DPO objective is

\begin{equation}
\begin{aligned}
\mathcal{L}_{\mathrm{DPO}}(\theta)
=
- \mathbb{E}_{(x,y_w,y_l)\sim \mathcal{D}}
\Bigg[
\log
\sigma
\Bigg(
\beta
\Bigg(
\log \frac{\pi_\theta(y_w \mid x)}
{\pi_{\mathrm{ref}}(y_w \mid x)}
\\
\qquad
-
\log \frac{\pi_\theta(y_l \mid x)}
{\pi_{\mathrm{ref}}(y_l \mid x)}
\Bigg)
\Bigg)
\Bigg]
\end{aligned}
\end{equation}
where $\pi_\theta$ denotes the current policy model, $\pi_{\mathrm{ref}}$ is the reference model, $\beta$ controls the strength of the reference constraint, and $\sigma$ is the sigmoid function.

Compared with conventional RLHF, DPO is easier to implement and often more stable to train. Nevertheless, its objective explicitly depends on the probability distribution of the reference model, and the update magnitude is controlled by $\beta$. When $\beta$ is large or preference data are noisy, the policy may move too far away from the reference distribution, resulting in degraded generation quality or mode collapse. This can be viewed as over-optimization in the DPO framework: once the reference model fails to serve as a reliable anchor, policy updates may continue in a biased direction.
	
\subsection{Iterative DPO and Over-Optimization}

To mitigate distribution shift, iterative DPO updates both the policy and reference models across multiple training rounds. Let $\pi_t$ denote the policy model at iteration $t$, which is typically used as the reference model for the next iteration:
 
\begin{equation}
\pi_{\mathrm{ref}}^{(t+1)} = \pi_t .
\end{equation}
The objective at iteration $t+1$ is then
\begin{equation}
\mathcal{L}_{\mathrm{DPO}}^{(t+1)} 
= 
\mathcal{L}_{\mathrm{DPO}}
\left(
\pi_{t+1} ; \pi_{\mathrm{ref}}^{(t+1)}
\right) .
\end{equation}
Therefore, the optimization objective changes whenever the reference model changes. If $\pi_t$ has been affected by noisy preferences, it may deviate from the true human preference distribution. Once it is used as the next reference model, this bias is propagated into the following objective.

Because the DPO loss contains reference-model log-probability terms, fluctuations in the reference model can alter the optimization direction and create non-stationary training dynamics. As iterations proceed, such errors may be progressively amplified, worsening over-optimization and causing performance oscillation. Thus, in iterative DPO, over-optimization is not only related to the choice of $\beta$ but also to the instability of the reference model.

\subsection{Model Merging}

A single model under continuous optimization can become overly specialized in a particular direction. Model merging provides a form of parameter-space ensembling that integrates the strengths of multiple models into a single checkpoint without increasing inference cost. Prior studies show that merged models can improve accuracy, robustness, and stability compared with individual models~\cite{fu2025model,cong2023have,feng2025aimmerging}. Common merging strategies include simple average and weighted average~\cite{lu2024merge}.

\subsubsection{Simple Average}
Simple average merging constructs a stronger model by averaging the parameters of multiple models. Given $\kappa$ models, the merged parameters are
\begin{equation}
\theta_{merge} =\frac{1}{\kappa } \sum_{i=1}^{\kappa } \theta_{i}  ,
\end{equation}
where $\kappa$ is the number of models and $\theta_i$ denotes the parameters of the $i$-th model.

\subsubsection{Weighted Average}
Compared with simple averaging, weighted averaging fusion can yield superior performance in the merged model~\cite{singh2020model}. This approach assigns different coefficients to each model according to its relative importance or quality, resulting in a more effective merged model. Therefore, determining appropriate weights for each model is a key factor in the merging process. The parameters of the resulting model after weighted averaging can be expressed as follows:

\begin{equation}
\theta_{merge} = \sum_{i=1}^{\kappa } \alpha_{i}  \theta_{i} ,
\end{equation}
where $\alpha_i$ is the normalized weight for the $i$-th model. Choosing appropriate weights is therefore central to the effectiveness of weighted model merging.

    \section{Method}
The overall iterative DPO training framework is illustrated in Figure~\ref{fig:mergedpo}.

\subsection{Motivation}

In iterative DPO, the policy model is progressively updated across multiple iterations, forming a continuous optimization trajectory. As the iterations proceed, the model can gradually improve its preference alignment capability, but it also becomes increasingly susceptible to preference noise and over-optimization, which may lead to performance degradation in the later stages of training.

Meanwhile, policy models from different training stages often exhibit complementary characteristics. Early-stage models tend to preserve stronger language modeling capabilities, intermediate-stage models achieve a better balance between alignment and stability, while late-stage models may suffer from performance deterioration. Motivated by these observations, we propose a learnable model merging approach based on historical trajectories. By performing weighted merging over policy models from different stages, the proposed method mitigates noise accumulation and alleviates the over-optimization issue in the later stages of iterative DPO training.

\subsection{Trajectory Model Merging with Learnable Weights}

To exploit policy information from different stages of iterative preference optimization, we consider the policy trajectory produced by iterative DPO:
\begin{equation}
\Theta =
\left\{
\theta^{(0)},
\theta^{(1)},
\dots,
\theta^{(T)}
\right\},
\end{equation}
where $\theta^{(t)}$ denotes the parameters of the policy model obtained after preference optimization at iteration $t$.

Instead of using only the previous-round policy as the reference model, we merge historical policies in parameter space to construct a trajectory-merged reference model for the next DPO iteration:
\begin{equation}
\theta^{*}
=
\sum_{t=0}^{T}
\alpha_t \theta^{(t)},
\end{equation}
where $\alpha_t$ is the merging weight assigned to the $t$-th historical policy.

To ensure that the merging weights satisfy probability constraints, we use a softmax parameterization:
\begin{equation}
\alpha = \mathrm{softmax}(w),
\quad
w \in \mathbb{R}^{T+1},
\end{equation}
which gives
\begin{equation}
\alpha_t \ge 0,
\quad
\sum_{t=0}^{T}\alpha_t = 1.
\end{equation}

During optimization, all historical policy parameters are frozen and only the weight parameter $w$ is updated. Given a preference dataset $\mathcal{D}=\{(x,y_w,y_l)\}$, we learn the merging weights with the following preference objective:

\begin{equation}
\mathcal{L}_{\mathrm{pref}}
=
-
\mathbb{E}_{(x,y_w,y_l)\sim\mathcal{D}}
\left[
\log
\sigma
\left(
\beta
\left(
\log \pi_{\theta^*}(y_w|x)
-
\log \pi_{\theta^*}(y_l|x)
\right)
\right)
\right] .
\end{equation}

Here $y_w$ and $y_l$ denote the preferred and rejected responses, respectively, and $\beta$ is the temperature coefficient. This objective encourages the merged model to assign a larger probability advantage to preferred responses, so that the learned weights favor historical policies with stronger preference alignment.

To prevent the weights from collapsing onto a single model, we add an entropy regularizer:

\begin{equation}
\mathcal{H}(\alpha)
=
-
\sum_{t=0}^{T}
\alpha_t \log \alpha_t .
\end{equation}

The final objective is

\begin{equation}
\mathcal{L}
=
\mathcal{L}_{\mathrm{pref}}
-
\lambda \mathcal{H}(\alpha),
\end{equation}

where $\lambda$ controls the strength of entropy regularization.

By learning adaptive weights over historical policies, TPMM-DPO integrates information from different optimization stages without retraining the historical model parameters. The resulting merged model provides a more stable reference for the next DPO iteration, reducing noise accumulation and late-stage over-optimization while improving the robustness of the final policy.

\section{Related Work}

\subsection{RLHF and Direct Preference Optimization}

Reinforcement Learning from Human Feedback (RLHF)~\cite{ouyang2022training} has become a mainstream paradigm for aligning large language models. It typically trains a reward model to capture human preferences and then optimizes the policy with reinforcement learning methods such as PPO. Although RLHF has been successful in practice, it is computationally expensive and depends heavily on stable reward modeling.

Direct Preference Optimization (DPO)~\cite{rafailov2023direct} was proposed to reduce the complexity of RLHF. DPO optimizes a log-odds loss constructed from preference comparisons and is theoretically connected to reward maximization with KL regularization. Owing to its simplicity and efficiency, DPO has become an important direction in preference alignment. Online and iterative variants of DPO~\cite{guo2024direct,pang2024iterative} use language models as annotators: in each iteration, the current model generates candidate responses, and a judge selects preferred and rejected samples for the next round of optimization. However, as model scale and the number of training iterations increase, iterative DPO can become unstable, especially when preference noise is present or when the reference model is repeatedly updated. This instability often appears as over-optimization~\cite{rafailov2024scaling}.

\subsection{Over-Optimization and Noise in DPO}

DPO assumes that preference data follow the Bradley--Terry--Luce model, but real human annotations often contain ambiguity and noise. When preference labels are incorrect or inconsistent, the model may overfit the training data, leading to degraded validation performance and weaker alignment. Tu et al.~\cite{tu2025enhancing} show that model performance decreases noticeably as preference noise increases. This issue is particularly pronounced in multi-round iterative DPO, where the reference model is updated at each iteration and noise-induced errors can accumulate along the training trajectory.

Existing work addresses this problem from several directions. One line of work strengthens KL regularization or introduces theoretical constraints to limit policy drift from the reference model~\cite{liu2024provably,nguyen2025mitigating}. These methods can help control policy divergence, but they do not fully address systematic noise in preference data.

Another line of work explicitly models preference noise. rDPO~\cite{chowdhury2024provably} constructs a debiased loss based on label-flip rates and provides consistency guarantees under random preference noise. Khaki et al.~\cite{khaki2024rs} use rejection sampling to construct more reliable preference pairs, while Phuc et al.~\cite{phucmitigating} reduce training variance through truncated importance ratios. The SPA framework~\cite{kim2024spread} mitigates annotation noise using confidence estimation and loss smoothing. These methods mainly focus on noise in single-round training. AIPO~\cite{shen2024aipo} combines an agreement-aware $\alpha$-DPO loss with NLL regularization, dynamically reweights samples using reference-model agreement scores, and filters low-quality conflicting samples to improve the stability of iterative preference optimization.

A third direction studies reference-model update strategies in DPO. MPO~\cite{gou2024mixed} uses a staged training mechanism. Kim et al.~\cite{kim2025sdpo} propose a simple but effective stepwise method, sDPO, which partitions the preference dataset and performs iterative DPO over these partitions while using the current policy as the reference model for the next iteration. Gorbatovski et al.~\cite{gorbatovski2024learn} propose TR-DPO, which dynamically updates the reference policy during training to address over-optimization in offline alignment. It combines soft updates, which interpolate the current and reference policies, with hard updates, which periodically replace the reference policy. These methods improve reference-model stability, but they do not explicitly model error accumulation over the policy trajectory.

\subsection{Model Merging and Parameter Averaging}

Model merging is an important technique for improving stability and generalization, and it has been widely used in ensemble learning and large-scale model training. Classical methods include simple averaging and weighted averaging, which merge parameters from multiple models to reduce the bias and overfitting of any single model~\cite{lu2024merge}. Cong et al.~\cite{cong2023have} use parameter averaging and related merging methods to study robustness in intellectual-property protection for large language models. Fu et al.~\cite{fu2025model} apply model merging to knowledge editing by combining model parameters containing new knowledge while preserving existing capabilities. Feng et al.~\cite{feng2025aimmerging} propose an adaptive iterative merging method based on training trajectories, dynamically integrating historical model parameters to improve stability, robustness, and knowledge retention in continual learning. These studies suggest that model merging can improve the stability and robustness of neural models.

	\section{Experimental Setup}
	\subsection{Datasets}
We use the helpful subset of the Anthropic Helpful and Harmless (HH) dataset~\cite{Training2022Bai} as the main training and evaluation benchmark. HH is a widely used pairwise preference dataset for alignment research. It contains 169K human preference dialogue pairs with prompt, chosen response, and rejected response fields, making it suitable for SFT, DPO, and preference-ranking evaluation~\cite{wu2024beta}. To evaluate cross-domain generalization, we additionally sample 1K examples from UltraFeedback~\cite{cui2023ultrafeedback} as an out-of-domain test set. UltraFeedback contains prompts collected from sources such as UltraChat and ShareGPT. These prompts are used to query multiple large language models, producing four responses per prompt and 256K responses in total.

	\subsection{Data Processing}
	HH provides pairwise preference data. We use chosen responses to construct the SFT training set and use $(\text{chosen}, \text{rejected})$ pairs as DPO training examples. To evaluate robustness under noisy preference data, we follow the setting of Tu et al.~\cite{tu2025enhancing} and apply random label-flip noise during DPO training. Specifically, for a flip probability $p\in[0,1]$, each preference pair swaps its chosen and rejected responses with probability $p$. Here $p=0.5$ means that half of the samples are flipped, while $p=1.0$ means that all preference labels are reversed.
	
	\subsection{Models}
	We evaluate our method using Llama3.2-3B~\cite{grattafiori2024llama} as the base model. During evaluation, UltraRM-13B~\cite{cui2023ultrafeedback} is used as the reward evaluator, denoted as the Gold RM, to assign scalar reward scores to generated responses. 

	\subsection{Training Pipeline}
	\paragraph{SFT} We first sample 10K examples from HH helpful-base for supervised fine-tuning and obtain the initial policy model. SFT uses full-parameter fine-tuning for 3 epochs, with a learning rate of $5\mathrm{e}{-6}$, batch size 16, maximum sequence length 1024, the Adam optimizer, \texttt{adam\_beta1}=0.9, \texttt{adam\_beta2}=0.95, and 600 warmup steps.

	\paragraph{DPO} In the DPO stage, we use 12K training examples from the HH helpful subset for preference optimization and sample 1K examples from the HH test subset as the main test set. The batch size is 8, and each iteration is trained for 1 epoch. For each batch, the model computes log probabilities of chosen and rejected responses under both the policy and reference models, and updates parameters using the DPO loss. We set $\beta=0.1$, use AdamW with a learning rate of $5\mathrm{e}{-7}$, and apply cosine annealing for learning-rate scheduling.

	In the iterative DPO framework, we run 3 rounds of optimization. After each round, the current policy generates candidate responses on the training set, and the Gold RM scores them to form new preference pairs. We then apply random label flips according to the target noise ratio and use the resulting data for the next DPO iteration. TPMM-DPO merges the current and historical policy models with learnable weights and uses the merged model as the reference model for the next DPO iteration. After all iterations, the policy model obtained in the final iteration is used as the final model.

	\subsection{Results}
	This section reports the main results, noise robustness, and iteration-wise win-rate analysis of the proposed learnable trajectory merging method. Unless otherwise specified, results are reported under the noise ratio $p=0.1$.

\subsubsection{Effect of the Entropy Regularization Coefficient}
We study the effect of the entropy coefficient $\lambda$, which controls the smoothness of the merging-weight distribution. As shown in Table~\ref{tab:lambda_ablation}, when $\lambda=0$, the learned weights tend to collapse onto a single checkpoint. This makes merging close to model selection and limits performance gains. As $\lambda$ increases, the weight distribution becomes smoother, allowing the model to better exploit complementary information from different stages. This leads to improvements on both HH and UF. However, overly large $\lambda$ values make the weights nearly uniform, weakening the ability to distinguish checkpoints of different quality and causing a mild performance drop. Overall, $\lambda=0.1$ provides the best balance between performance and stability, and we use it as the default setting.
\begin{table}[htbp]
\caption{Effect of entropy regularization coefficient $\lambda$ on Llama3.2-3B.}
\label{tab:lambda_ablation}
\centering
\small
\begin{tabular}{lcc}
\toprule
Model & Win Rate (\%) & OOD Win Rate (\%) \\
\midrule
TPMM-DPO($\lambda$=0.0)   & 63.0 $\pm$ 0.5 & 57.1 $\pm$ 0.5 \\
TPMM-DPO($\lambda$=0.01)  & 65.2 $\pm$ 0.4 & 57.6 $\pm$ 0.6 \\
TPMM-DPO($\lambda$=0.05)  & 67.1 $\pm$ 0.3 & 58.2 $\pm$ 0.4 \\
TPMM-DPO($\lambda$=0.1)   & \textbf{68.4 $\pm$ 0.4} & \textbf{58.7 $\pm$ 0.7} \\
TPMM-DPO($\lambda$=0.2)   & 67.0 $\pm$ 0.6 & 58.1 $\pm$ 0.5 \\
TPMM-DPO($\lambda$=0.5)   & 64.8 $\pm$ 0.7 & 57.3 $\pm$ 0.6 \\
\bottomrule
\end{tabular}
\end{table}

\subsubsection{Overall Performance Comparison}

We evaluate TPMM-DPO on Llama3.2-3B and compare it with SFT, iterative DPO, sDPO, and rDPO. The results are shown in Table~\ref{tab:main_results_llama}. All win rates are computed by GPT-4.1-nano as a pairwise judge, using the preference rate over SFT outputs as the evaluation criterion. Gold Reward is computed using UltraRM-13B as the reward model.
Overall, iterative DPO shows a non-monotonic trend across multiple training iterations. Its performance peaks at the second iteration and then drops at the third iteration, suggesting that later optimization can introduce performance degradation and weaker generalization. Using a simple-average merged model as the next reference model reduces the late-stage drop, but its win rate remains lower than that of the learnable weighted merging method.

TPMM-DPO achieves the best results across all evaluation metrics. At the third iteration, its win rate reaches 68.4\%, clearly improving over the best single-round DPO checkpoint. It also shows stronger out-of-domain generalization, with an OOD win rate of 58.7\%.

In addition, TPMM-DPO outperforms sDPO and rDPO in win rate, indicating that trajectory-aware learnable parameter merging can more effectively integrate policy information from different optimization stages and yield more stable final performance.

\begin{table*}[htbp]
\centering
\small
\caption{Overall performance comparison on Llama3.2-3B.}
\label{tab:main_results_llama}
\begin{tabular}{lcccc}
\toprule
Model & Win Rate & Gold Reward & OOD Win Rate & OOD Gold Reward \\
\midrule

SFT & 50.0 & 0.582 & 50.0 & 0.105 \\

\midrule
DPO Iter1 & 56.8 & 0.613 & 54.2 & 0.214 \\
DPO Iter2 & 61.5 & 0.621 & 57.4 & 0.238 \\
DPO Iter3 & 59.7 & 0.589 & 51.1 & 0.201 \\

\midrule
sDPO & 61.8 & 0.627 & 55.3 & 0.226 \\
rDPO & 65.9 & 0.664 & 57.9 & 0.271 \\

\midrule
TPMM-DPO\\
(Simple Average) Iter2 & 62.4 & 0.643 & 57.0 & 0.234 \\
TPMM-DPO\\
(Simple Average) Iter3 & 63.7 & 0.629 & 57.2 & 0.266 \\

TPMM-DPO(Ours) Iter2 & 62.9 & 0.671 & 57.6 & 0.242 \\
TPMM-DPO(Ours) Iter3 & \textbf{68.4} & \textbf{0.681} & \textbf{58.7} & \textbf{0.287} \\

\bottomrule
\end{tabular}
\end{table*}

\subsubsection{Robustness to Preference Noise}

Table~\ref{tab:noise} reports win rates under different noise ratios $p$. As expected, performance generally decreases as the noise level increases, showing that corrupted preference signals directly weaken alignment.

For standard iterative DPO, performance improves across iterations in the noise-free setting ($p=0.0$), increasing from 58.2 to 63.9. Under noisy settings, however, later checkpoints degrade noticeably. For example, when $p=0.7$, Iter3 drops from 49.2 at Iter1 to 42.8, indicating that noise can be accumulated and amplified through multi-round optimization.

By contrast, TPMM-DPO shows a more stable trend. Since both methods use the SFT model as the reference model in the first iteration, their Iter1 results are identical. Starting from the second iteration, TPMM-DPO brings clear improvements under moderate noise ($p=0.3$), reaching 58.9 compared with 53.2 for standard DPO. This suggests that cross-iteration model merging can effectively reduce error accumulation caused by noisy preferences. Under higher noise levels ($p=0.5,0.7$), although the overall performance is lower, TPMM-DPO remains comparable to or slightly better than the corresponding DPO checkpoints, demonstrating stronger robustness.

\begin{figure}[htbp]
\centering
\begin{subfigure}{0.45\textwidth}
\includegraphics[width=\linewidth]{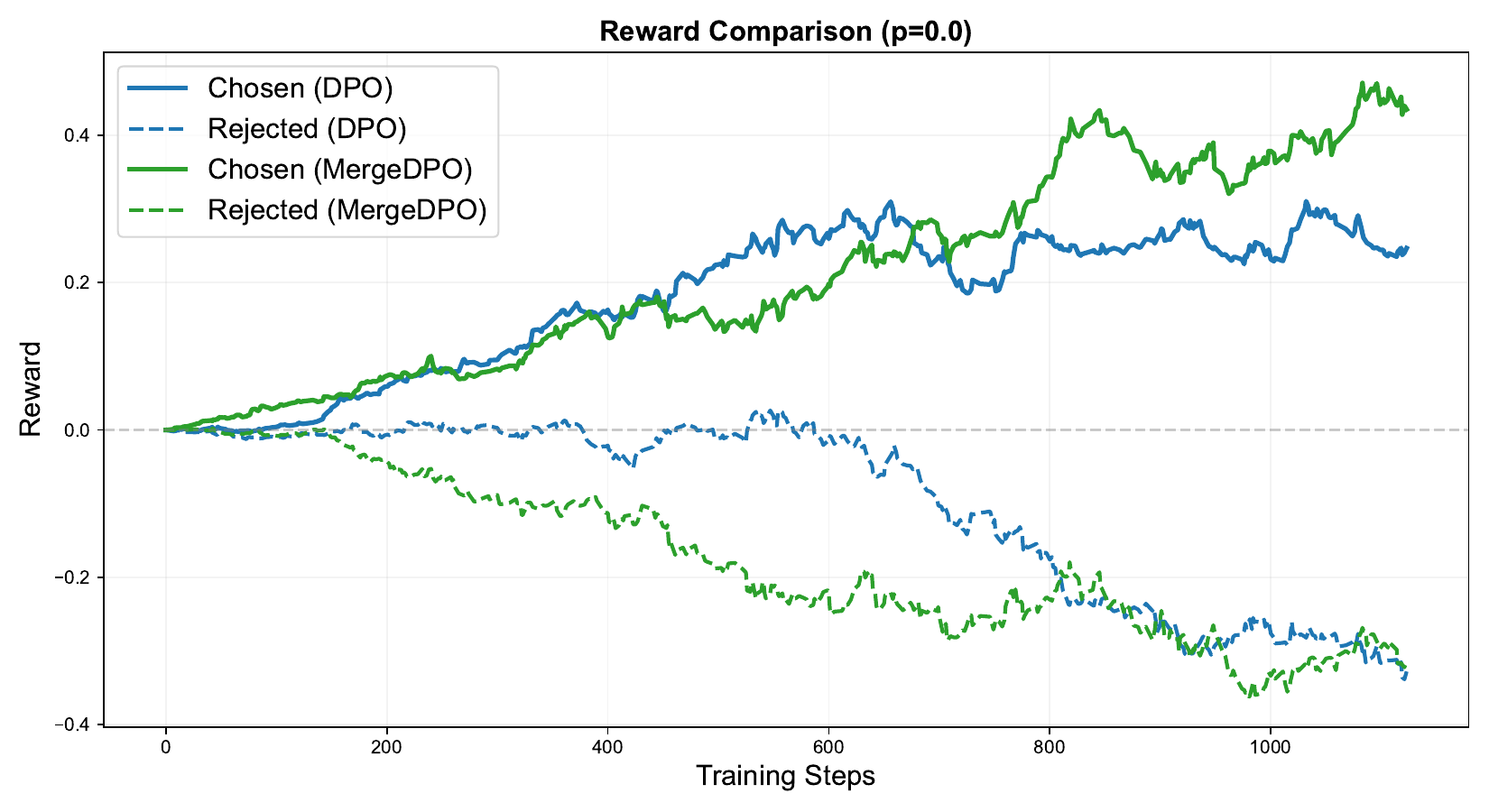}
\end{subfigure}
\begin{subfigure}{0.45\textwidth}
\includegraphics[width=\linewidth]{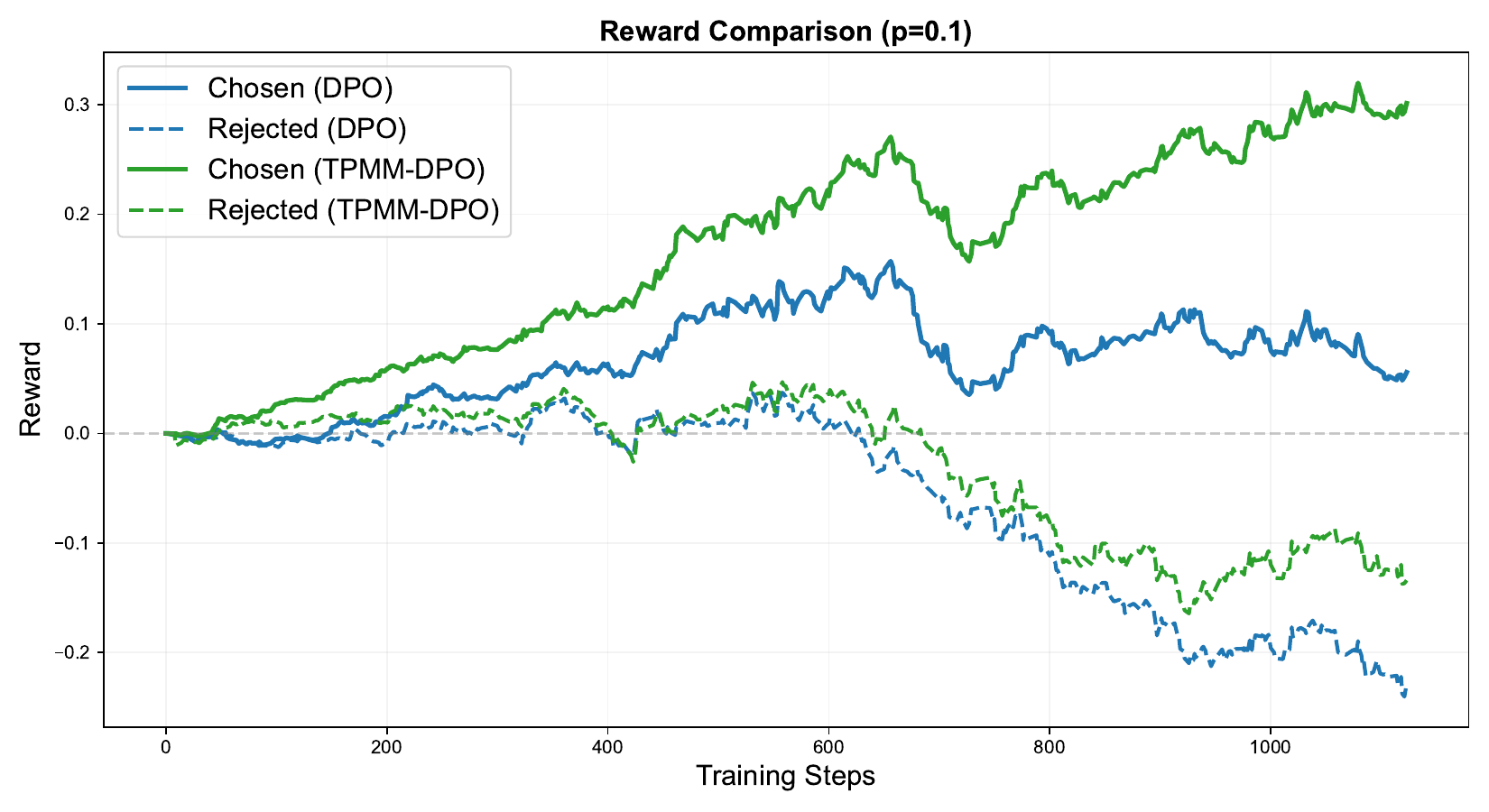}
\end{subfigure}

\vspace{0.5em}

\begin{subfigure}{0.45\textwidth}
\includegraphics[width=\linewidth]{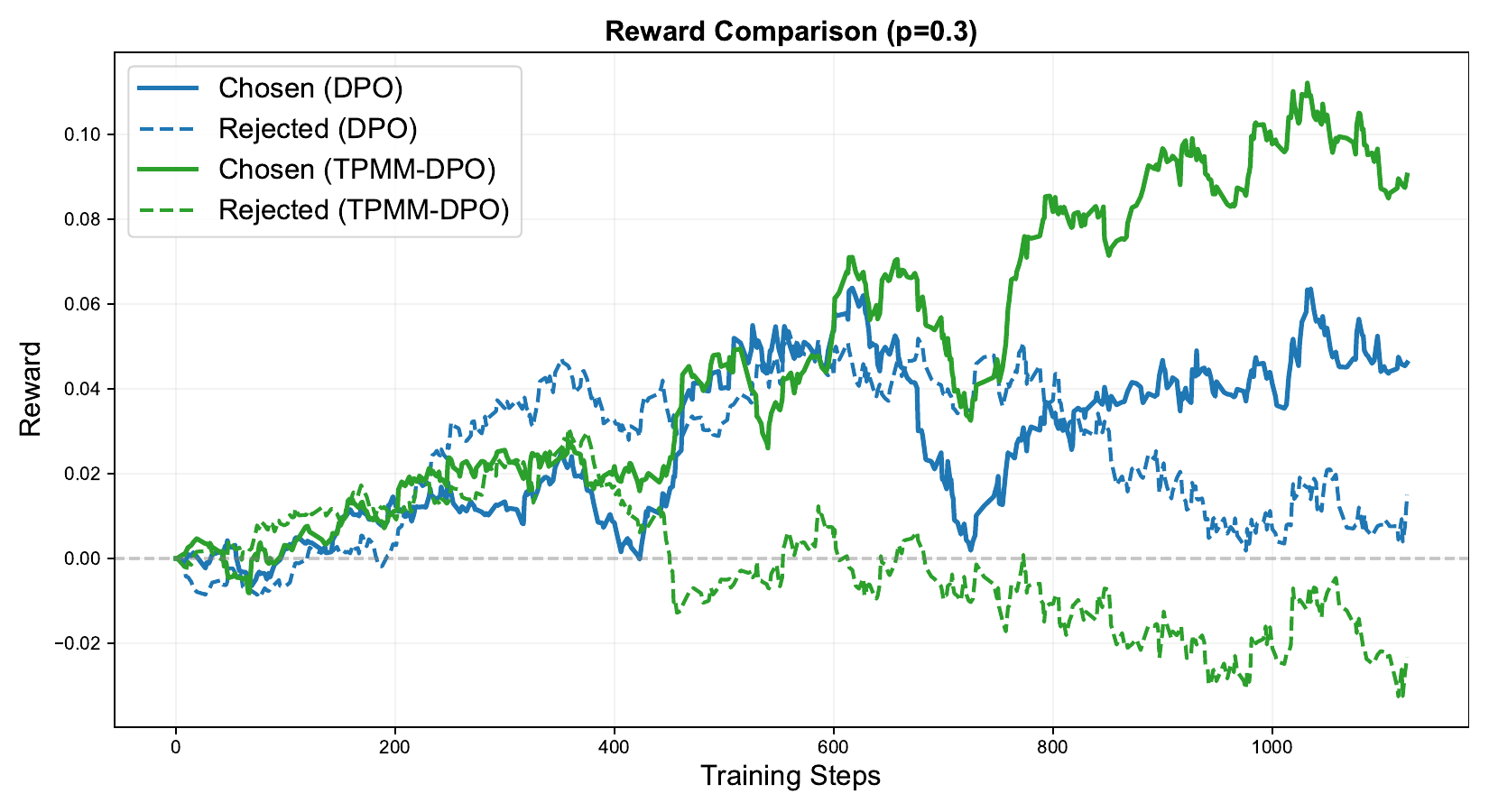}
\end{subfigure}
\begin{subfigure}{0.45\textwidth}
\includegraphics[width=\linewidth]{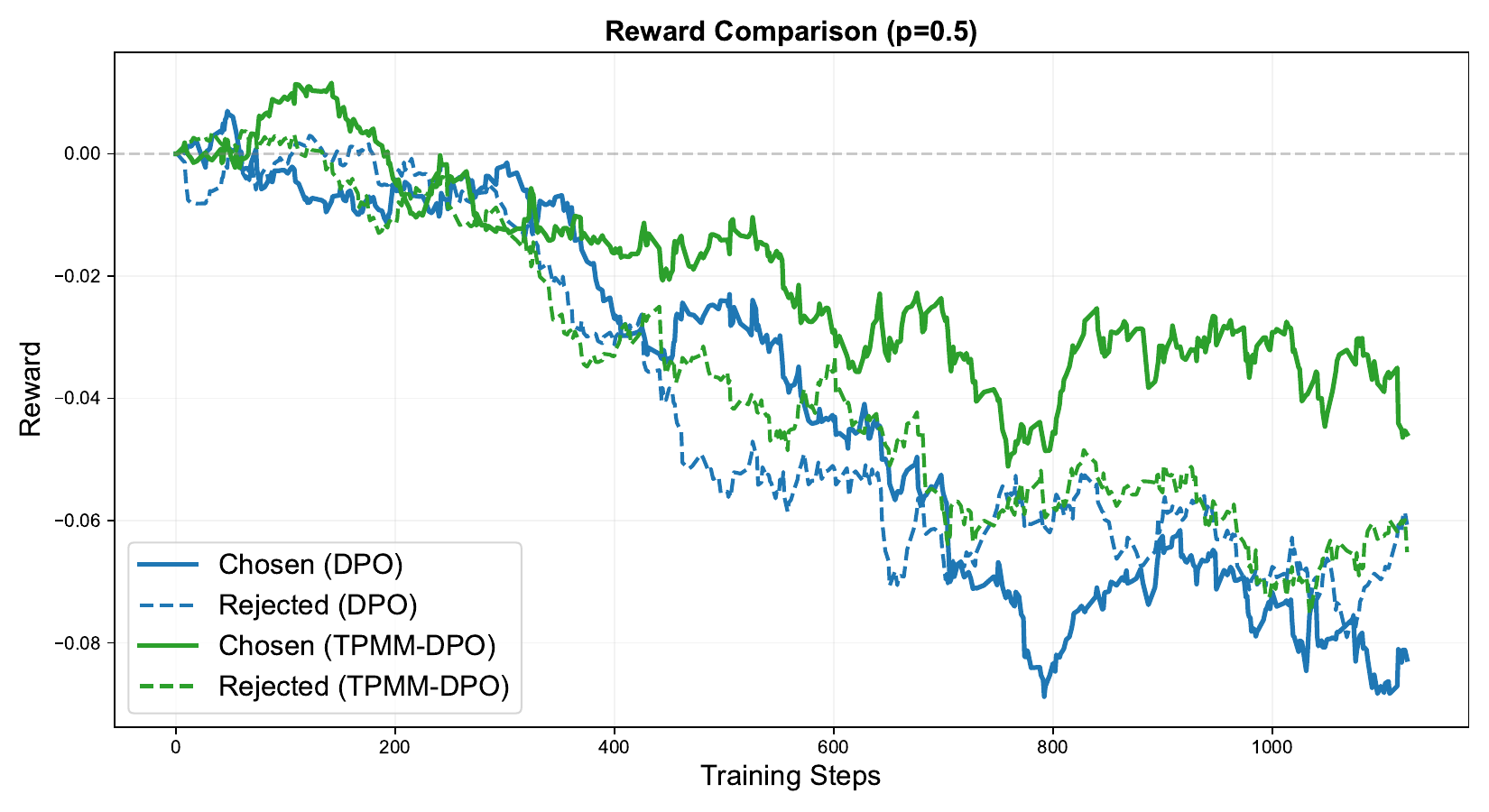}
\end{subfigure}

\vspace{0.5em}

\begin{subfigure}{0.45\textwidth}
\includegraphics[width=\linewidth]{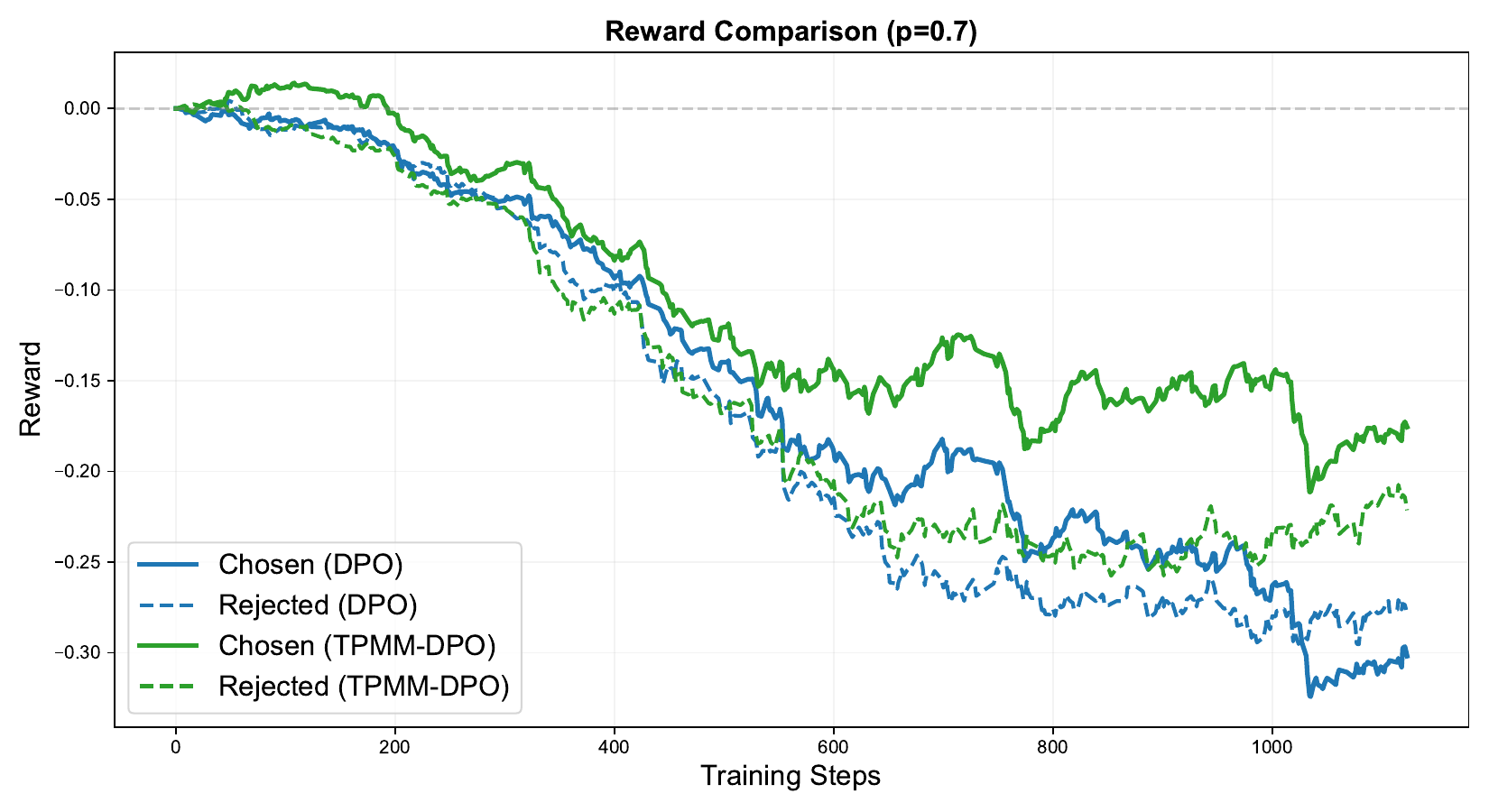}
\end{subfigure}

\caption{Comparison of the chosen and rejected rewards after three rounds of iterative training for iterative DPO and TPMM-DPO under different noise ratios. Solid lines represent the chosen rewards, while dashed lines represent the rejected rewards.}
\label{fig:noise-reward}
\end{figure}

To further examine robustness, we compare the reward curves of iterative DPO and TPMM-DPO under $p\in\{0.0,0.1,0.3,0.5,0.7\}$, as shown in Figure~\ref{fig:noise-reward}. Under low noise ($p\in\{0.0,0.1\}$), both methods gradually increase the reward gap between chosen and rejected responses, while TPMM-DPO converges faster and separates them more clearly. Under moderate noise ($p=0.3$), standard DPO exhibits strong oscillation in reward differences, whereas TPMM-DPO continues to enlarge the positive-negative reward margin. Under high noise ($p\in\{0.5,0.7\}$), standard DPO is more strongly affected by noisy preferences and shows a contracted reward gap, while TPMM-DPO maintains a larger separation and is less sensitive to noise.

Overall, TPMM-DPO effectively mitigates performance degradation in noisy iterative DPO and provides notable gains under moderate noise.

\begin{table}[htbp]
\caption{Robustness to preference noise on HH with different reference model strategies (Llama3.2-3B). TPMM-DPO uses learned weights to fuse historical models as the reference model.}
\label{tab:noise}
\centering
\small
\begin{tabular}{lccccc}
\toprule
Model & $p=0.0$ & $p=0.1$ & $p=0.3$ & $p=0.5$ & $p=0.7$ \\
\midrule
DPO Iter1 & 58.2 &56.8 & 53.0 & 50.6 & \textbf{49.2} \\
DPO Iter2 & 62.0  &61.5 & 55.4 & 50.1 & 46.7 \\
DPO Iter3 & 63.9 &59.7 & 53.2 & 49.0 & 42.8 \\
\midrule
TPMM-DPO Iter1 & 58.2 &56.8 & 53.0 & 50.6 & \textbf{49.2} \\
TPMM-DPO Iter2 & 66.1 &62.9 & 57.4 & \textbf{51.3} & 48.5 \\
TPMM-DPO Iter3 & \textbf{70.8} & \textbf{68.4} & \textbf{58.9} & 50.1 & 46.4 \\
\bottomrule
\end{tabular}
\end{table}

\subsubsection{Performance Trends Across Iterations}

\begin{figure}[htbp]
    \centering
    \includegraphics[width=0.45\linewidth]{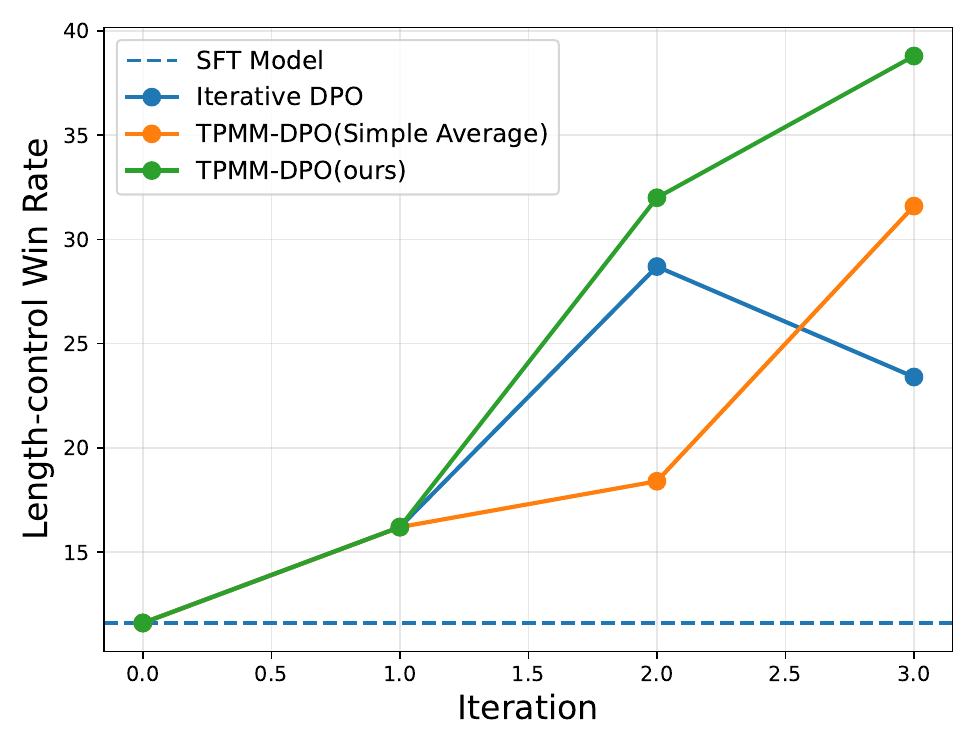}
    \caption{Length-controlled win-rate comparison across iterative training rounds.}
    \label{fig:iteration_curve}
\end{figure}

Figure~\ref{fig:iteration_curve} illustrates the variation trends of the length-control win rate across multiple training iterations for different methods. This metric is obtained by comparing the generated responses with those produced by GPT-4.1-nano and is used to evaluate the generation quality of the models.

Iterative DPO reaches its peak performance at the second iteration (28.7) but declines significantly to 23.4 in the third iteration, indicating the existence of performance degradation during the later stages of training. TPMM-DPO (Simple Average), which adopts the simple average of historical models as the reference model, achieves relatively stable performance improvements and reaches 31.6 in the third iteration.

In contrast, the proposed TPMM-DPO learns adaptive weights for different historical models and performs weighted fusion to construct the reference model for the next iteration, thereby enabling continuous performance improvement throughout the iterative process. Starting from the second iteration, TPMM-DPO (ours) already demonstrates clear advantages over the other methods and further achieves the highest score of 38.8 in the third iteration, substantially outperforming both Iterative DPO and TPMM-DPO (Simple Average).

Overall, TPMM-DPO (ours) effectively improves generation quality during iterative training and alleviates the performance degradation issue encountered by standard Iterative DPO in the later stages.

	\section{Conclusions}
	
This paper addresses noise accumulation and late-stage over-optimization in multi-round preference optimization under iterative DPO. We find that using only the previous iteration’s policy model as the reference can suffer from training instability, length bias, and degraded generalization in later stages. To mitigate this, we view historical policy models from iterative DPO as an optimization trajectory and propose a trajectory-based fusion method with learnable weights. Specifically, we construct a weighted combination of models from different iterations and learn the fusion coefficients using a preference optimization objective, enabling adaptive integration of complementary knowledge across stages and yielding a more stable reference model for subsequent DPO training.

Unlike conventional iterative DPO that relies on a single historical checkpoint as the reference, our method avoids retraining model parameters and instead optimizes a small set of fusion weights to merge historical trajectories, thereby reducing performance degradation from late-stage over-optimization with minimal computational overhead. Experiments on noisy preference data show that while individual checkpoints tend to degrade in middle and late training stages, the proposed trajectory-fused reference model improves training stability and achieves better final performance and generalization.

Future work may explore more fine-grained dynamic weighting strategies, such as sample- or uncertainty-aware fusion, and extend the method to larger-scale models and more complex scenarios, including safety alignment, factual consistency, and long-form reasoning, to further assess its generality and practical effectiveness.
	


\end{document}